\documentclass[a4paper,UKenglish,cleveref, autoref, thm-restate]{lipics-v2021}

\pdfoutput=1
\hideLIPIcs  

\title{Baba is You is Undecidable} 

\author{Jonathan Geller}{Department of Computer Science, Williams College, Williamstown, Massachusetts, United States}{jmg8@williams.edu}{https://orcid.org/0000-0002-7631-6146 }{}

\authorrunning{J. Geller} 

\Copyright{Jonathan Geller}

\ccsdesc[300]{Theory of computation~Problems, reductions and completeness}

\keywords{Baba is You, Undecidability, Turing-Completeness} 

\acknowledgements{The author would like to thank Aaron Williams for many helpful discussions on the topic.}

\nolinenumbers 

\newcommand{\rl}[1]{\textsc{#1}}

\begin{document}
\maketitle 

\begin{abstract}
  
  We establish the undecidability of 2019 puzzle game \textit{Baba is You} through a reduction from the Post correspondence problem. In particular, we consider a restricted form of the Post correspondence problem introduced by Neary (STACS 2015) that is limited to five pairs of words. \textit{Baba is You} is an award winning tile-based game in which the player can reprogram the game's mechanisms by pushing blocks that spell out the rules. We achieve undecidability through a generalization of the size of the playfield in the horizontal direction, adding a ``hallway'' to one side of the level. The undecidability of Baba is You has been claimed several times online using different source problems, including the simulation of Turing machines and Conway's Game of Life, however, this contribution appears to be the first formal proof of the result.

\end{abstract}

\section{Introduction}

Many puzzles and games have been shown to admit reductions from NP- or PSPACE- hard problems, though it is rare that solutions to such puzzles cannot be found in PSPACE. While some video games, particularly Dwarf Fortress, Cities: Skylines \cite{bali_2019}, and Minecraft, have been shown informally to be Turing-Complete, the author is only aware of more formal proofs of the undecidability of Braid \cite{hamilton2014braid} and Magic: the Gathering \cite{DBLP:journals/corr/abs-1904-09828}. The computational complexity of sliding block puzzles such as Sokoban and Rush Hour, however, has been studied extensively. A general framework for demonstrating PSPACE-completeness of such puzzles is presented in \cite{DBLP:journals/corr/cs-CC-0205005}.

\begin{figure}
  \begin{center}
    \includegraphics[scale=0.2]{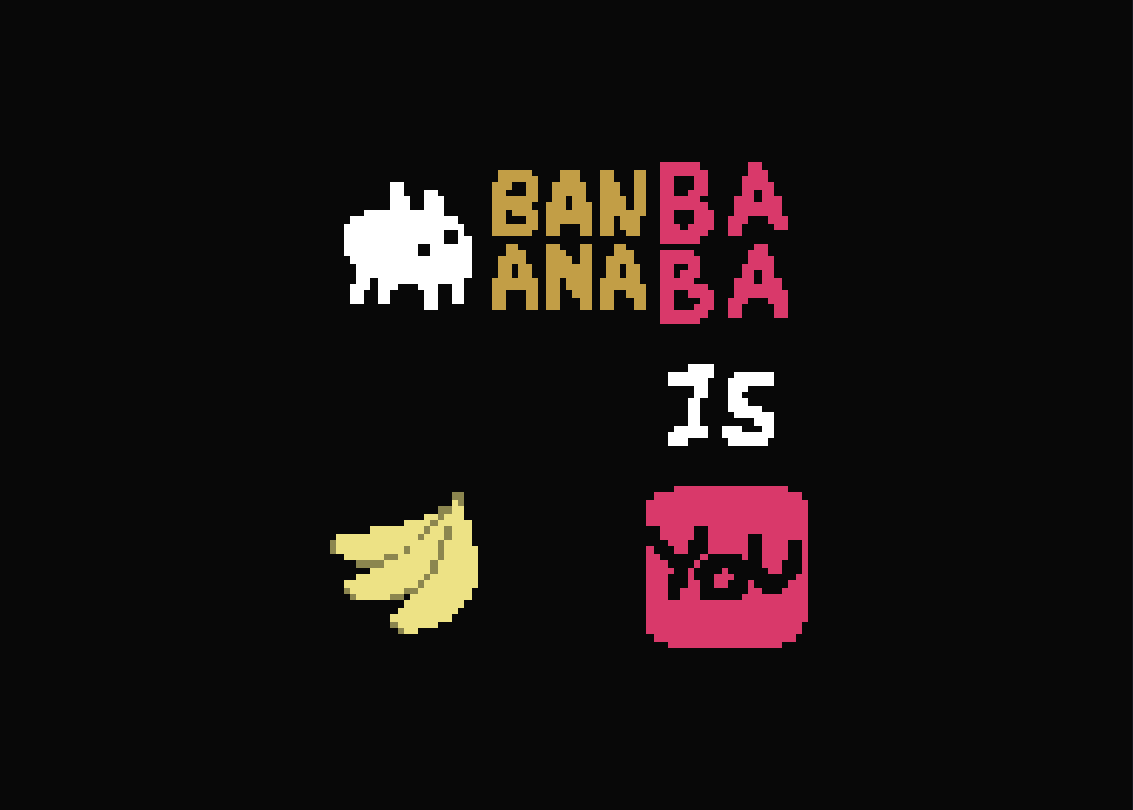}
    \includegraphics[scale=0.2]{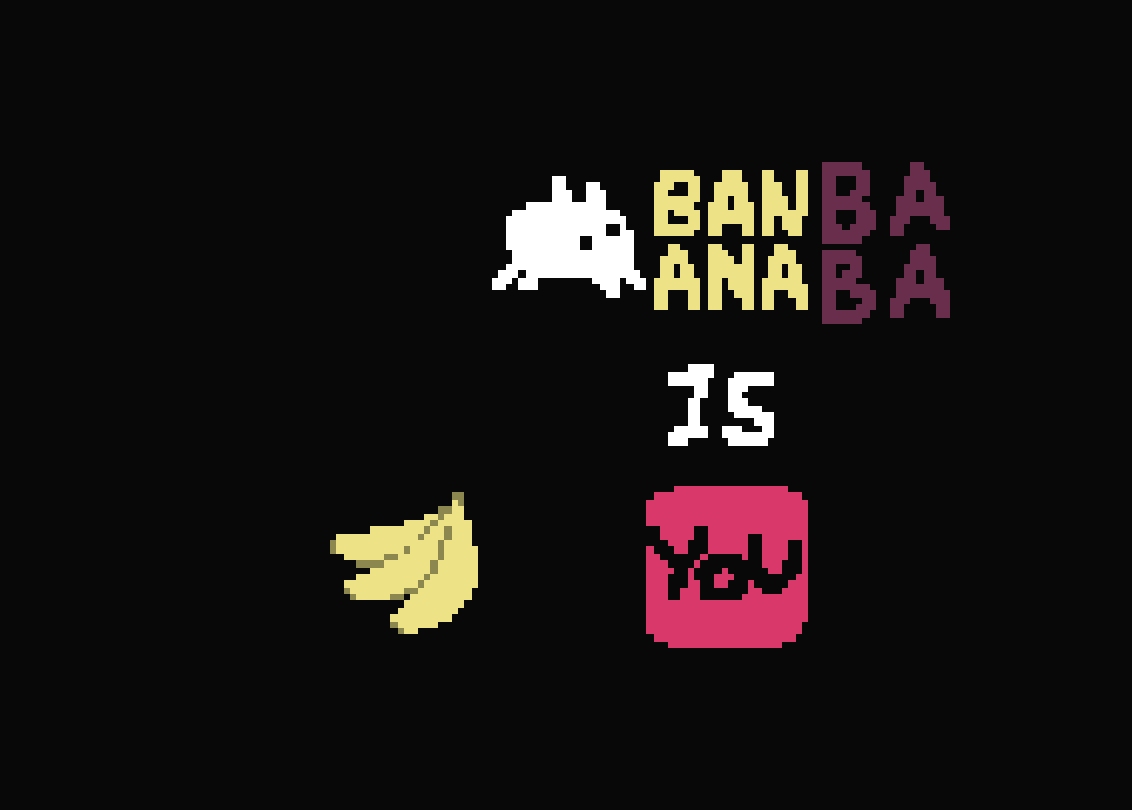}
    \includegraphics[scale=0.2]{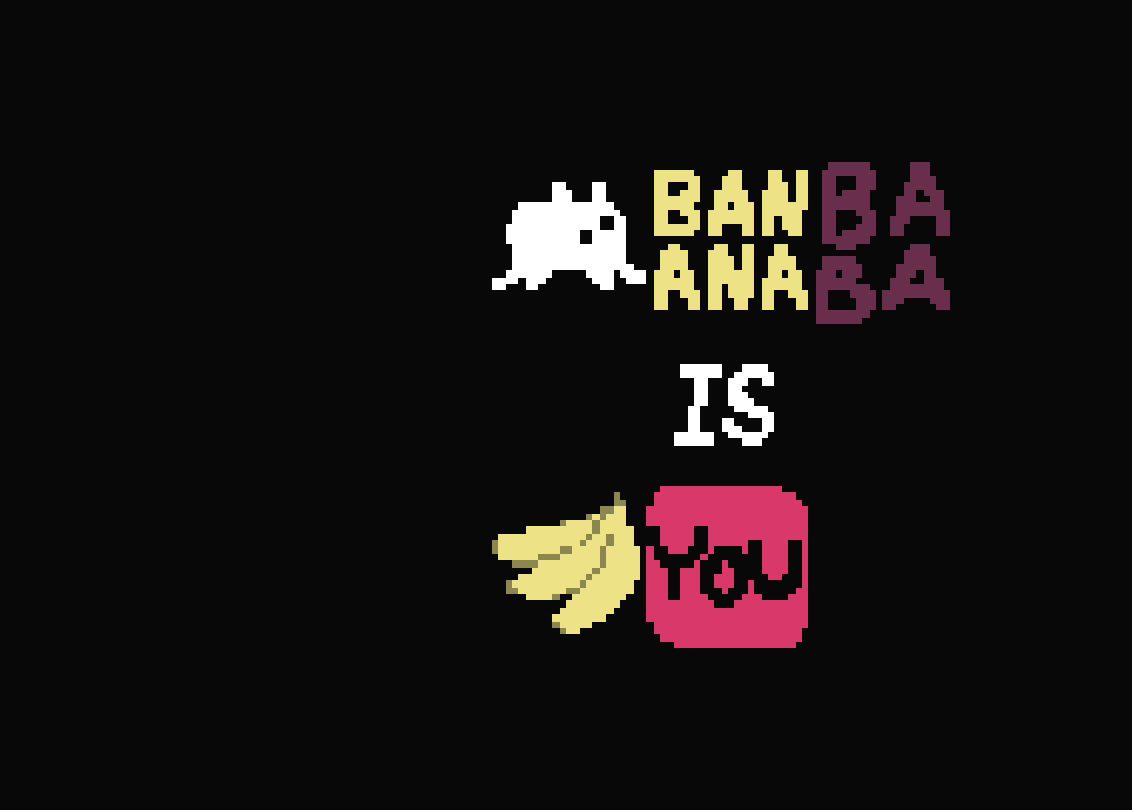}
  \end{center}
  \caption{An example section of a \textit{Baba is You} level. In the first image, the rule \rl{baba is you} means that if the player moves to the right, the baba object will move to the right, pushing the text \rl{banana} and \rl{baba} to the right, as seen in the second image. Now, the rule formed is \rl{banana is you}, so if the player moves right again, the banana object will move rather than the baba object. }\label{fig:demo}
\end{figure}

The 2019 puzzle game \textit{Baba is You} \cite{teikari2019baba} could be classified as a sliding block puzzle, but we will see that the game admits much more complex behavior than almost any other. Levels generally consist of players pushing objects around to reach some sort of end goal, much like any other block-pushing puzzle game. Unlike these other games, however, \textit{Baba is You} allows the player to constantly reshape the rules of the game, as rules are formatted as sentences which exist in the level as interactable objects. No objects innately have special properties: all properties of an object are granted by rules which exist in the level as text, as is visible in Figure~\ref{fig:demo}. For a more detailed explanation of how other aspects of the game used in this reduction work, see the videos in \cite{babaVideos} which contain brief demonstrations of all relevant game mechanics.

Using the conventional restrictions on levels in the game results in a bounded number of reachable states from any initial finite level, so the game with no modifications is decidable. However, we present here a slight modification to the function of the level boundary which enables a reduction from the Post correspondence problem to the game \textit{Baba is You}. In particular, this is apparently the first undecidability result for any block-pushing puzzle game.

\begin{definition}
  An instance of \textsc{baba+2} consists of some finite rectangular level with an additional 2-tile tall ``hallway'' appended to the right side of the level. Objects may pass arbitrarily far into this hallway while having their positions stored, and the top and bottom of this hallway behave just as the level border normally does. Such an instance is a yes instance if there is some finite sequence of moves which can be executed that guarantees that a player will at some point reach the \rl{win} victory condition, ignoring Too Complex and Infinite Loop end conditions in any level state with a valid interpretation of all rules, and without ever forming rules of the form \rl{empty is \_\_\_}.
\end{definition}

The reasoning behind the formalisms of this definition is presented in Section 2 along with more detailed information about \textit{Baba is You}, and a full reduction from a variant of the Post correspondence problem to \textsc{baba+2} is presented in Section 3.

\section{Background}

\subsection{More on \textit{Baba is You}}

As alluded to earlier, an unmodified version of \textit{Baba is You} is still decidable. If more than six copies of a single object are stacked in one location, the number of those objects on that tile is automatically reset to six. Additionally, there are only a finite number of distinct objects available in the game. Keeping these restrictions in place, we find that the number of states which are reachable from some initial level is bounded by some exponential in the size of the level. As a result, if none of these restrictions are relaxed, there does exist an algorithm that can determine if a given \textit{Baba is You} level can be won by simply calculating the result of every possible move at every possible state reachable from some initial level and searching for a sequence of moves that results in a victory.

A few examples of tools which could be used to prove Turing-completeness of a modified version of \textit{Baba is You} have been demonstrated, such as the construction of Conway's \textit{Game of Life} in \cite{gameOfLife1} and \cite{gameOfLife2}, though none of these examples have explicitly stated what generalizations to \textit{Baba is You} they require. Additionally, none of these constructions include demonstrations of how their gadgets can be used to win or lose a level, so the existence of these levels does not strictly rule out that there could be an algorithm which determines if a generalized \textit{Baba is You} level can be won. Here, we present a simple generalization of \textit{Baba is You} which can be the object of a formal reduction from a known undecidable problem.

\subsection{Formalizing the decision problems}

The ultimate goal of the reduction is to reduce a simplified version of the Post correspondence problem to a slight modification of \textit{Baba is You}. First, we define the decision problem \textsc{pcp5-2}:

\begin{definition}
  An instance of \textsc{pcp5-2} consists of ten binary strings, \(\alpha_1\), \(\alpha_2\), \(\alpha_3\), \(\alpha_4\), \(\alpha_5\), \(\beta_1\), \(\beta_2\), \(\beta_3\), \(\beta_4\), \(\beta_5\). An instance of \textsc{pcp5-2} is a yes instance when there exists some finite nonempty sequence of integers from 1 to 5, \((i_1,i_2,i_3,\cdots,i_n)\) such that \(\alpha_{i_1}\alpha_{i_2}\cdots\alpha_{i_n}=\beta_{i_1}\beta_{i_2}\cdots\beta_{i_n}\), where \(\alpha_1\alpha_2\) represents concatenation of the two strings. 
\end{definition}

The version of this problem in which the ten strings are constructed from an arbitrary finite alphabet has been shown to be undecidable in \cite{neary:LIPIcs:2015:4948}. It is easy to reduce that problem to \textsc{pcp5-2} by associating each character in an alphabet \(\Sigma\) with a distinct binary string of length \(\lceil \log_2|\Sigma|\rceil\) and replacing each string in the original instance of the problem with the corresponding binary string. Thus, \textsc{pcp5-2} is undecidable.

Defining an appropriate decision problem for \textit{Baba is You} is more complicated. A few aspects of the game in particular must be taken into consideration to ensure that our decision problem is well defined. When there are more than two copies of an object with the \rl{tele} property, the game selects where to teleport objects at random, and so undoing and redoing the same move can lead to different results. Additionally, if a certain recursion depth is exceeded in resolving the rules of a level at a particular timestep, or if certain aspects of the rules become too computationally difficult to resolve in any other way, the game will exit the level and display a screen with ``Infinite Loop'' or ``Too Complex,'' respectively. Since this is a limit imposed by the difficulty of computing rules of the game and not the rules of the game itself, we will ignore it. Embedding a genuine paradox in the rules still counts as losing the level, since there is no valid way to continue the level (consider the rule \rl{not group is group}, which has no valid resolution), but any set of rules that can be resolved will allow the level to continue, regardless of how computationally complex it is to evaluate the correct interpretation of the rules.

Finally, in order to demonstrate that a version of \textit{Baba is You} is undecidable, we will need some way of storing arbitrary amounts of information in a single level. A few possible methods exist for this, but here we will consider a modification in which an arbitrarily long, two-tile-tall space is appended to the edge of an otherwise finite level. This space begins empty, but if other tiles somehow enter this space, their positions are stored and they continue to behave as they would anywhere else in the level. The top and bottom of this additional space behave in the same fashion as a normal level border, and any objects/text must begin the level inside some finite-sized level. Finally, in the event that a rule such as \rl{empty is rock} is formed, the empty tiles in the hallway would not all be turned into rocks, resulting in the creation of infinitely many rocks, a situation with no clear resolution. Thus, we disallow forming rules of the form \rl{empty is}. This all gives rise to the definition of \textsc{baba+2} presented in Section 1.

\section{The reduction}

Consider the following instance of \textsc{pcp5-2}.

\begin{table*}[!ht]
    \begin{center}
      \begin{tabular}{|c|c|c|c|c|c|}
        \hline
        \(i\) & 1 & 2 & 3 & 4 & 5 \\
        \hline \hline
        \(\alpha_i\) & 1 & 0 & 010 & 11 & 010 \\ 
        \hline
        \(\beta_i\) & 10 & 10 & 01 & 1 & 10\\
        \hline
      \end{tabular}
    \end{center}
\end{table*}

This instance of \textsc{pcp5-2} is a yes instance, as demonstrated by the following solution:

\begin{table*}[!ht]
    \begin{center}
      \begin{tabular}{|c|c|c|c|c|c|c|}
        \hline
        \(i\) & 1 & 2 & 1 & 3 & 3 & 4 \\
        \hline \hline
        \(\alpha_i\) & 1 & 0 & 1 & 010 & 010 & 11 \\ 
        \hline
        \(\beta_i\) & 10 & 10 & 10 & 01 & 01 & 1\\
        \hline
      \end{tabular}
    \end{center}
\end{table*}

We demonstrate a particular \textsc{baba+2} instance corresponding to this instance of \textsc{pcp5-2}, then generalize to show how a level can be constructed in general to be solvable if and only if a given instance of \textsc{pcp5-2} is also solvable.

\subsection{Writing strings}

\begin{figure}[h!]
    \begin{center}
    \includegraphics[width=5.5in]{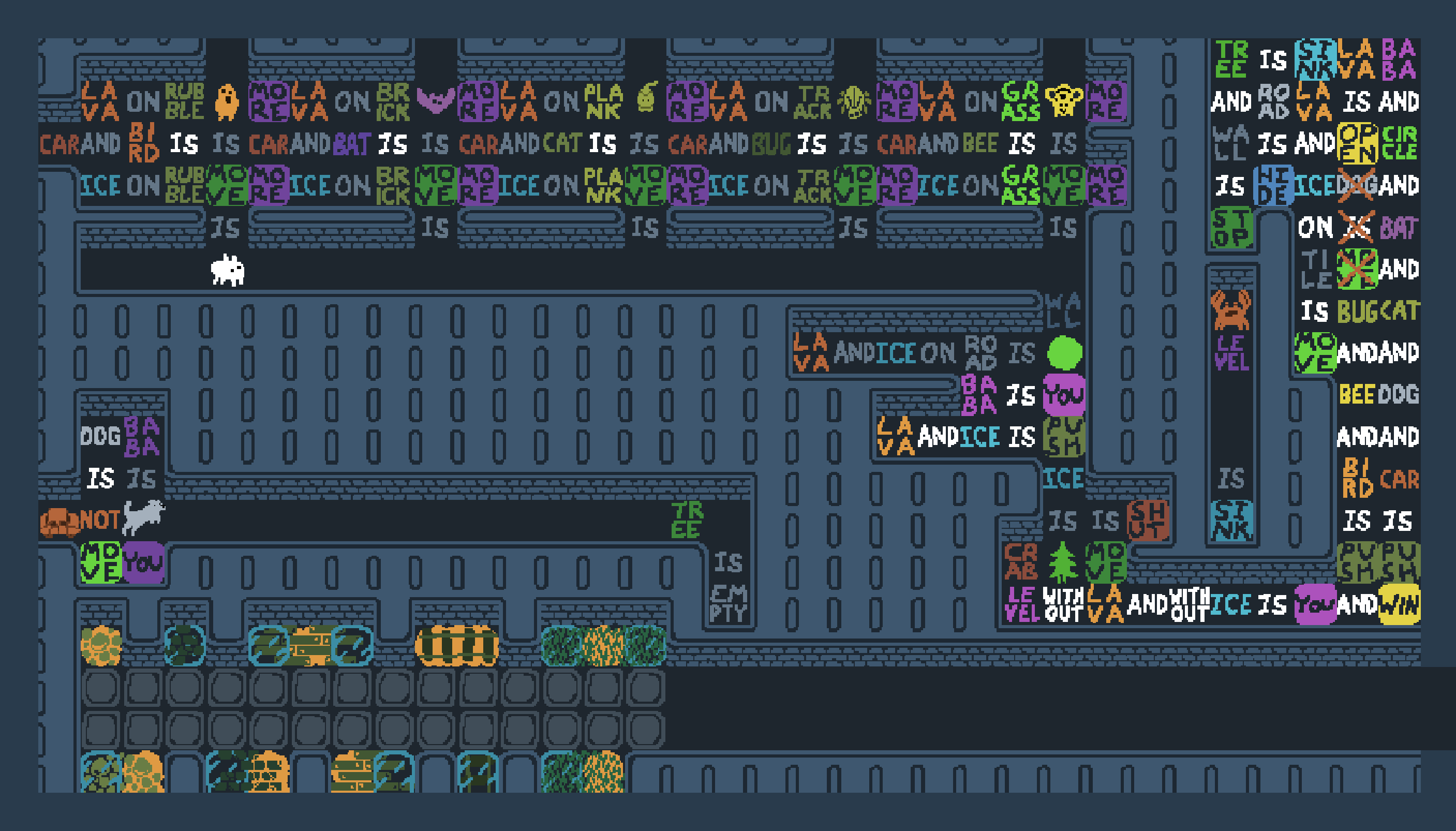}
    \end{center}
    \caption[]{Example instance of \textsc{baba+2} constructed from the earlier instance of \textsc{pcp5-2}. Road objects are present on top of each lava and ice object, but are hidden to allow the other objects on those tiles to be visible. All lava and ice objects are right-facing. Note the positioning of the "hallway" at the bottom right of the level.}\label{fig:full}

\end{figure}

Figure~\ref{fig:full} depicts an instance of \textsc{baba+2} corresponding to the above \textsc{pcp5-2} instance. It certainly should not be clear at this point how that problem corresponds to this level, though this image will serve as a reference for the remainder of this section.

In any \textsc{baba+2} instance we create, there are five buttons that will be present, which when pressed will perform an action equivalent to making a copy of one of the tiles from the \textsc{pcp5-2} instance. 

\begin{figure}[h!]
    \begin{center}
    \includegraphics[height=1in]{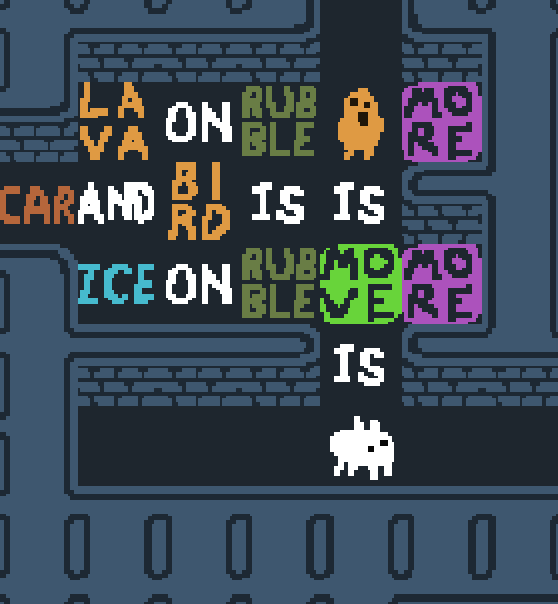}
    \includegraphics[height=1in]{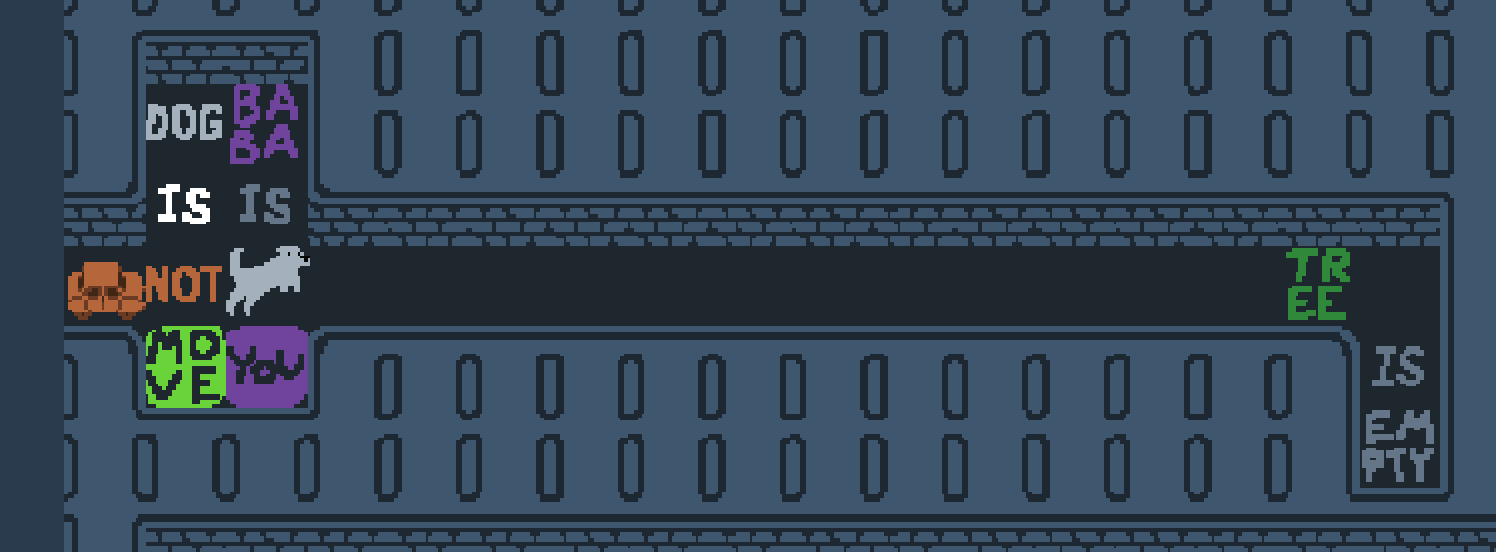}
    \includegraphics[height=1in]{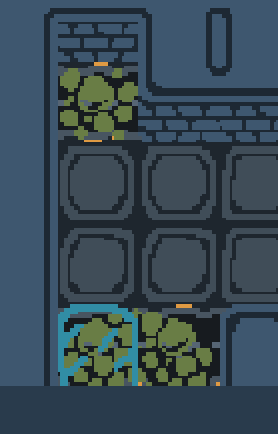}
    \end{center}
    \caption[]{Three components of the level: from left to right, a button gadget, the timer gadget, and a duplication gadget. The duplication gadget corresponds to the tile with \(\alpha =1\), \(\beta=10\) from the earlier instance of \textsc{pcp5-2}. Important rules found elsewhere in the level are \rl{wall is stop}, \rl{baba is you}, \rl{lava and ice on tile is move}, \rl{bird is push}, \rl{car is push}, \rl{baba is push}, \rl{lava and ice is push}, and \rl{dog is move}.}\label{fig:gadgets}

\end{figure}

\begin{figure}[h!]
    \begin{subfigure}{.48\textwidth}
      \centering
      \includegraphics[height=1in]{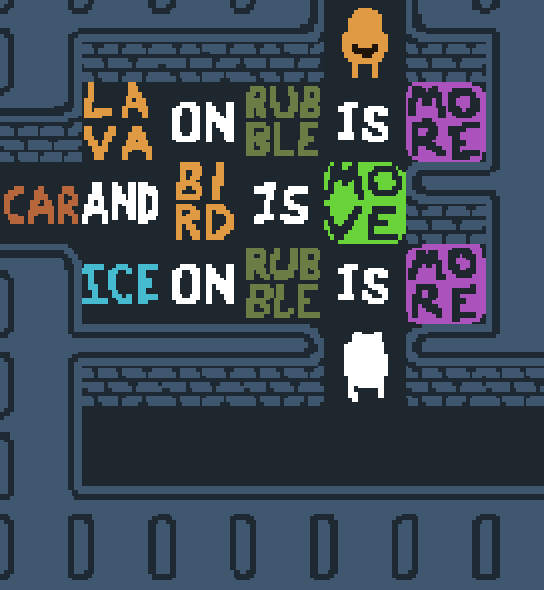}
      \includegraphics[height=1in]{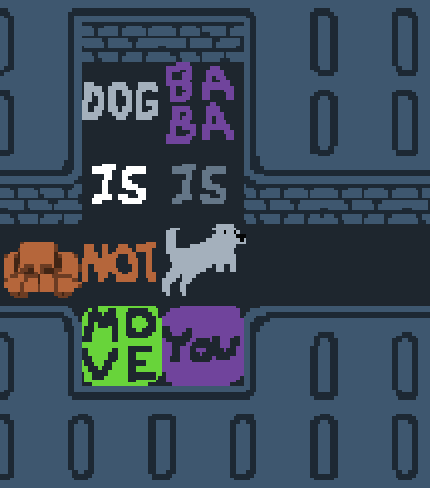}
      \includegraphics[height=1in]{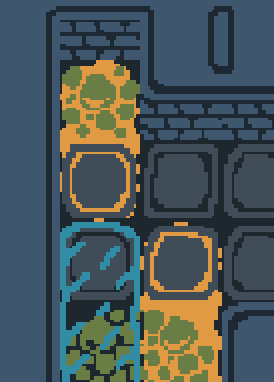}
      \caption{The gadgets from Figure~\ref{fig:gadgets} after the player moves up, duplicating the lava and ice objects in the third image. Nothing in the timer gadget changes on this timestep.}
      \label{fig:sfig1}
    \end{subfigure}
    \begin{subfigure}{.04\textwidth}
    \end{subfigure}
    \begin{subfigure}{.48\textwidth}
      \centering
      \includegraphics[height=1in]{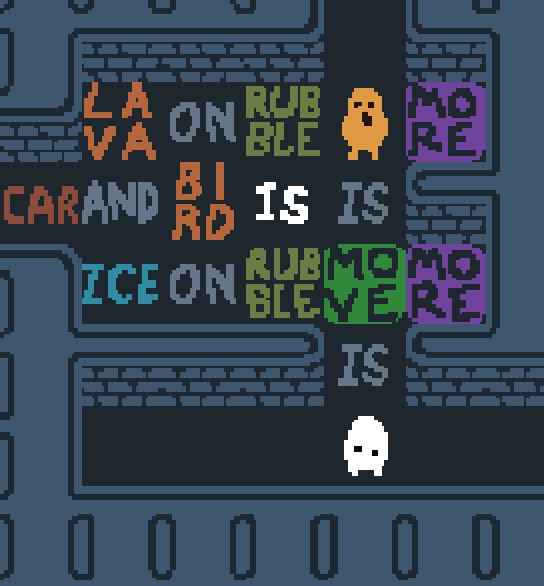}
      \includegraphics[height=1in]{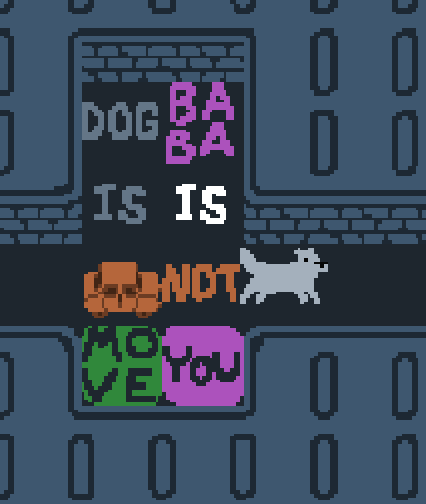}
      \includegraphics[height=1in]{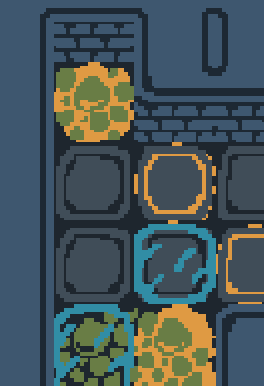}
      \caption{The same gadgets after one additional move in any direction, including waiting. Note that baba is pushed out of the first gadget, and the dog, lava, and ice objects begin to move right. }
      \label{fig:sfig2}
    \end{subfigure}
  \caption[]{If the player pushes up on the \rl{is} text shown in the first image of Figure~\ref{fig:gadgets}, each of the gadgets takes the state shown in (a). After this, the player's moves do not matter, and the state in (b) will be reached no matter what. }\label{fig:gadgets2}

\end{figure}

The button gadget displayed in the first image of Figure~\ref{fig:gadgets} is accessible to the baba object (which the player has control of), so the player may push the \rl{is} text upwards. This is the only way the player may interact with the text visible in the image. Doing so forms the rules \rl{ice on rubble is more}, \rl{lava on rubble is more}, duplicating the lava and ice objects in the third image of Figure~\ref{fig:gadgets}, as seen in Figure~\ref{fig:sfig1}. On the next time step, the new rule \rl{car and bird is move} causes the car in the second image to move to the right, breaking the rule \rl{dog is not move} and forming the rule \rl{baba is not you}, as seen in Figure~\ref{fig:sfig2}. The bird will also move downwards, pushing baba out of the gadget and resetting the gadget to the state it is in in the image above. This occurs in the exact same way no matter what action the player takes, since the player's move processes first, and moving right, left, or up will have no result.

In the following time steps, the dog in the second image will move all the way to the right, forming the rule \rl{tree is empty} if it has not yet been formed, though on subsequent iterations where this gadget is activated, the dog will still travel the same distance before turning around, bouncing off the rule it previously formed without modifying it. The dog will eventually travel back to the left, pushing the car back into place, reforming \rl{dog is not move}, and breaking \rl{baba is not you}, returning to the player control of baba. While this happens, the newly duplicated lava and ice objects will move to the right per the rule \rl{lava and ice on tile is move} until they reach the end of the strip of tile objects depicted at the bottom of the level in the full-level image. Once those objects reach the end of this strip, they will push other lava and ice objects already present so that they are further to the right, and they will stop just after the tile objects in the same left-to-right order as the order in which they were produced.

Note that no objects are \rl{you} during this entire process, so the player cannot interfere in any way. The strip that the dog object must move across is constructed to be at least as long as the strip of tile objects at the bottom of the level, so that the player must wait more turns than it requires for the lava and ice objects to move off of the tile objects before they may duplicate any more objects. This stops the player from duplicating two sets of objects in such a way that would allow them to add lava and ice objects in an order other than one of the ten strings already present at the beginning of the level.

We construct the arrangement of lava and ice objects in the third image of Figure~\ref{fig:gadgets} from the tile in the \textsc{pcp5-2} instance in the following fashion: we read the top string (1) in order from left to right, adding a lava tile in the top row for each 1 and an ice tile in the top row for each 0. We then read the bottom string (10), placing an ice tile in the bottom row for each 1, and a lava tile for each 0.

A copy of the gadgets shown in the first and third images of Figure~\ref{fig:gadgets} should be made for each tile in the \textsc{pcp5-2} instance, swapping the bird and rubble objects and text for other objects and their corresponding texts. In the full level image, the pairs are, in order, bird/rubble, bat/brick, cat/plank, bug/track, bee/grass. Only one copy of the gadget with the dog is needed, since only one of the buttons may be pressed at once, so the same gadget with the dog may be used as a timer for all of the buttons.

\subsection{Comparing strings}

\begin{figure}[h!]  
  \begin{center}
      \includegraphics[height=1.2in]{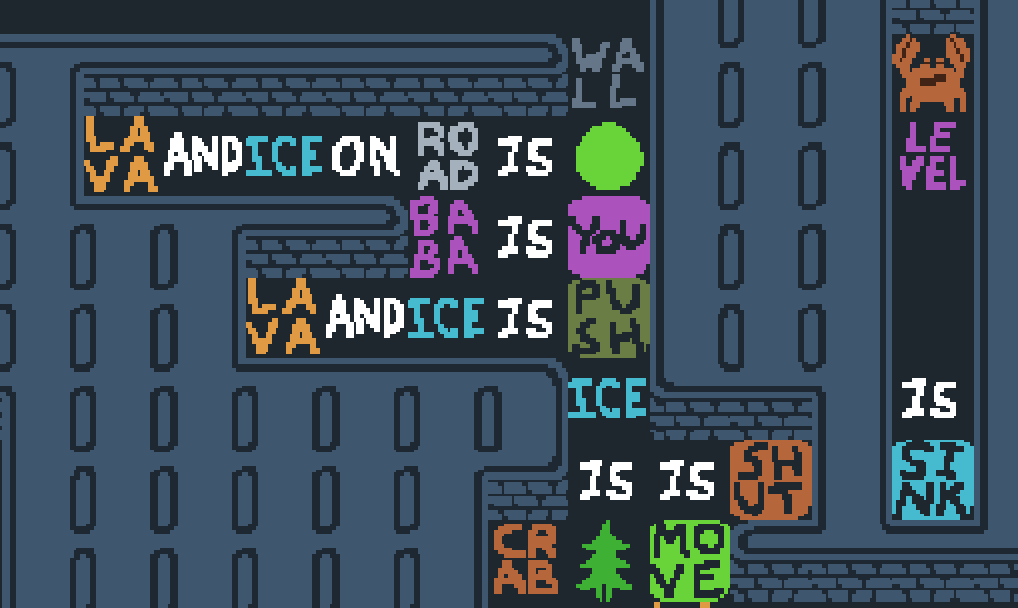}
  \end{center}
  \caption[]{The gadget with which the player can interact when they have created two matching strings. Pushing down on \rl{wall} enables the player to win the game if and only if they have produced strings of lava and ice in the top and bottom row of the hallway that correspond to matching binary strings. Additional relevant rules are \rl{lava is open}, \rl{circle is push}, \rl{tree is stop}, and \rl{level without lava and without ice is you and win}.}\label{fig:final}
\end{figure}

We may now discuss the section of the level visible at the center right in the image of the full level, or in Figure~\ref{fig:final}. The text \rl{wall} is accessible to the player. Because \rl{tree is stop}, the player may not push the column of ``\rl{wall} (circle) \rl{you push ice is}'' downwards. As seen earlier in the timer gadget, however, once at least one of the buttons has been pressed once, the rule \rl{tree is empty} will be formed, removing the tree from this column and enabling the player to push \rl{wall} downwards, forming and breaking various rules. In order for the player to push \rl{wall} downards, the player must have control of baba, which can only occur if all lava and ice objects at the bottom of the screen have moved to the right off of the tiles. When the player pushes \rl{wall} down, the rules \rl{baba is you} and \rl{lava and ice is push} are broken, and the rules \rl{crab is move}, \rl{ice is shut}, \rl{lava and ice is you}, and \rl{lava and ice on road is wall} are formed. This relinquishes control of baba from the player, locking off access from all of the button gadgets, and causes the crab to begin pushing \rl{level} downwards to form the rule \rl{level is sink}. This rule takes priority over any other actions, and renders the level unbeatable by deleting all objects and text on the screen. This will occur in two turns and cannot be stopped or delayed, meaning that after pushing \rl{wall} downwards, the player must win on the next turn or they will be unable to win.

The win condition for the level comes from \rl{level without lava and without ice is you and win} (no other win condition can be formed, since the \rl{win} word is inaccessible). The ice and lava objects present at the beginning of the level are all stacked on top of road objects, meaning that when the rule \rl{lava and ice on road is wall} is formed, they will be deleted. The only lava and ice objects remaining are then those which have been produced via the button gadgets and aligned  just to the right of the tile objects. There must be at least one such lava or ice object, since as discussed earlier, the player may only push \rl{wall} downwards if at least one button has been pressed already, meaning the player has duplicated at least one lava or ice object.

On the player's next turn, if they take the wait action, nothing will happen to the lava and ice tiles, and they will lose on the next turn no matter what. Similarly, if the player takes the right or left action, they will move all lava and ice tiles to the left or right, meaning they will still be present and the level will not be won.\footnote{Due to a strange way the game processes movement, it is necessary for the \rl{lava and ice is push} rule to be broken for this to function as intended–if the rule were still present then it would be possible to annihilate ice and lava objects by moving left or right. However, the rule is broken, so this does in fact work as intended.} Thus, the only option is for the player to move up or down. 

In the case of moving down, the lava and ice objects in the bottom of the two rows will fail to move since they are up against a wall, and the lava and ice objects in the top row will move down into the lava and ice objects in the bottom row. Moving up has the same effect, in which lava and ice objects in the same column will collide. Since \rl{ice is shut} and \rl{lava is open}, any column which contains one ice object and one lava object (in either order) will have those two objects annihilate each other via the \rl{open}/\rl{shut} rule. Thus, if in each column there is exactly one ice and one lava, then moving down or up will lead to no ice and lava remaining in the level, and so the rule \rl{level without lava and without ice is you and win} takes effect and causes the player to win the level. Alternatively, if there is a column which contains two ice, two lava, or a column with exactly one object (e.g. a single ice object in the top row with no matching object in the bottom row), then the objects will not be annihilated so at least one object of ice or lava will remain in the level, meaning the player will not win on that turn. Since the player will lose on the next turn regardless of what action is taken, the player will be unable to win.

\subsection{Constructing the level}

With the function of the individual components of the level clarified, we now present a general scheme for constructing a level based off of a general \textsc{pcp5-2} instance. It should be noted that this procedure will not produce levels with the exact same layout as the level in Figure~\ref{fig:full}, as that level in particular has been rearranged to be more compact for ease of presentation. Still, it should be clear how the level that the following procedure constructs is equivalent to the level already presented.

\begin{figure}[h!]
\begin{center}
    \includegraphics[height=2.4in]{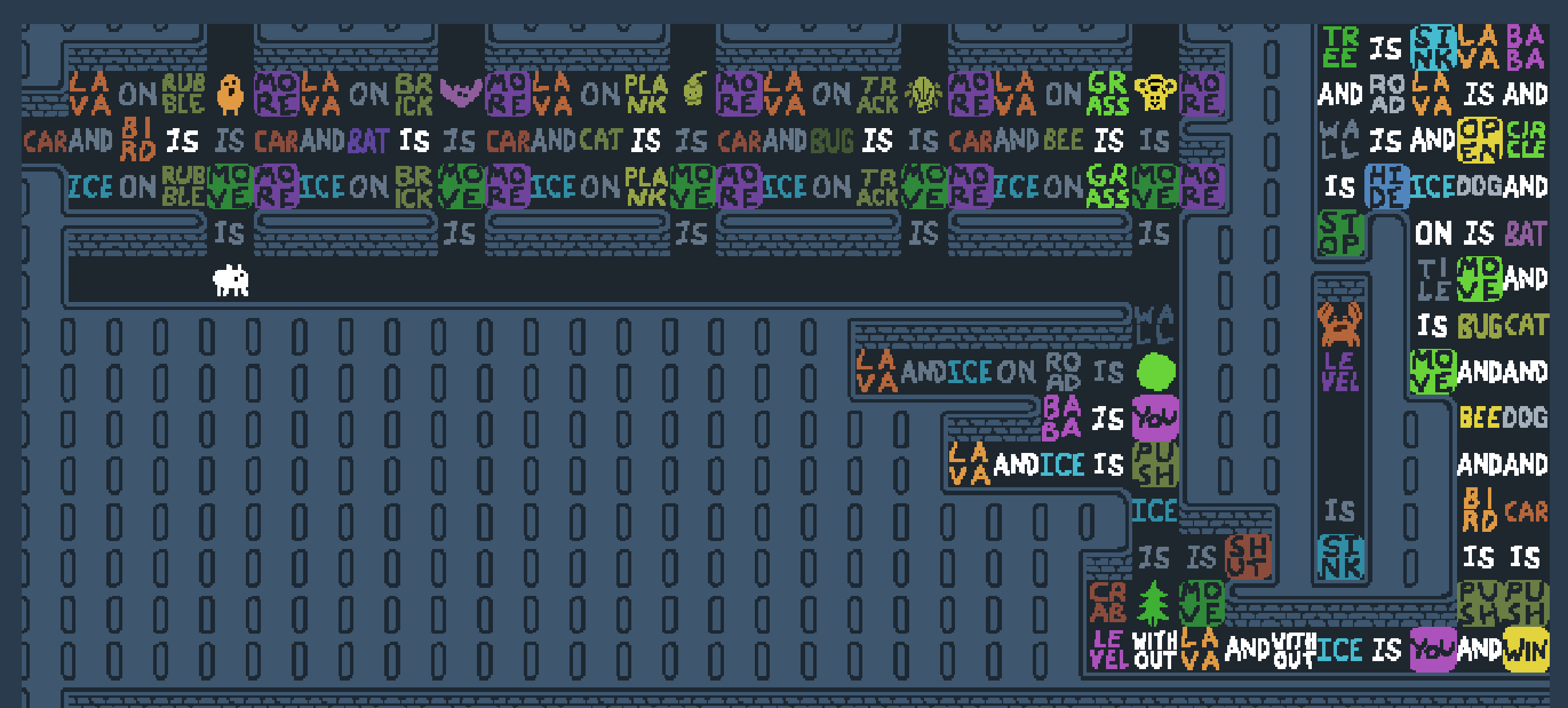}
\end{center}
\caption[]{The top section of every instance of \textsc{baba+2} created. The contents of this section are identical for any initial instance of \textsc{pcp5-2}. The right level border need not be where it is shown here, though any additional space remaining to the right of this will not be used.}\label{fig:top}
\end{figure}

\begin{figure}[h!]
\begin{center}
    \includegraphics[height=1in]{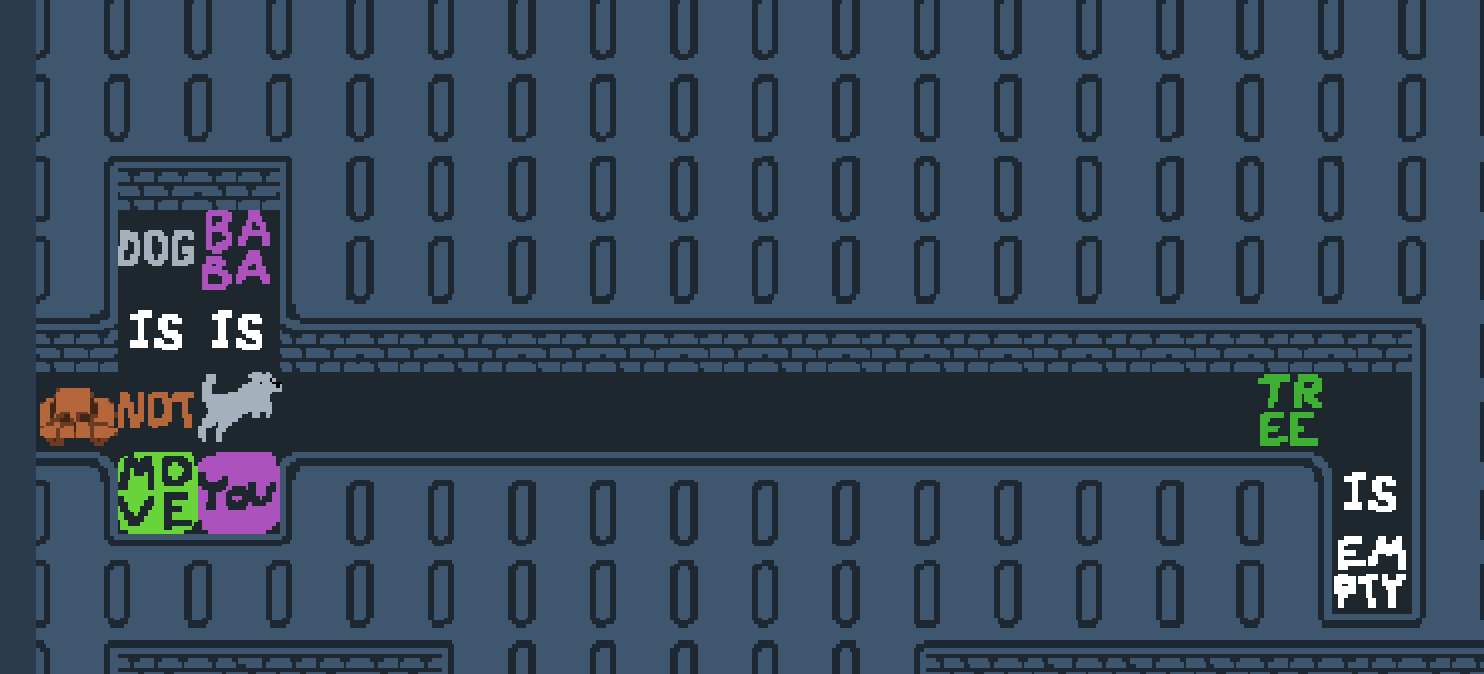}
\end{center}
\caption[]{The timer gadget to be constructed in each instance of \textsc{baba+2}. This should be modified to span the full width of the level in order to ensure that the timer is sufficiently long for each particular \textsc{baba+2} instance.}\label{fig:timer}  
\end{figure}

Given an instance of \textsc{pcp5-2}, there are ten binary strings \(\alpha_1\), \(\alpha_2\), \(\alpha_3\), \(\alpha_4\), \(\alpha_5\), \(\beta_1\), \(\beta_2\), \(\beta_3\), \(\beta_4\), and \(\beta_5\). Define the quantity \(l\):
\[l=\max\left\{\sum_{i=1}^5 |\alpha_i|,\sum_{i=1}^5 |\beta_i|,33\right\}\]
where \(|\alpha_i|\) denotes the length of the string \(\alpha_i\). We construct a level of width \(l+6\) and height 25 beginning with the level section seen in Figure~\ref{fig:top} aligned to the top left corner of the level. Subsequently, we construct the timer gadget of Figure~\ref{fig:timer} below the previous part, but stretched out so that the left and right edges of the timer line up with the left and right edges of the level.

Fill the fifth row from the bottom of the level with walls. Fill every space in the second and third rows from the bottom with tile objects. 

\begin{figure}[h!]
  \begin{center}
      \includegraphics[height=0.8in]{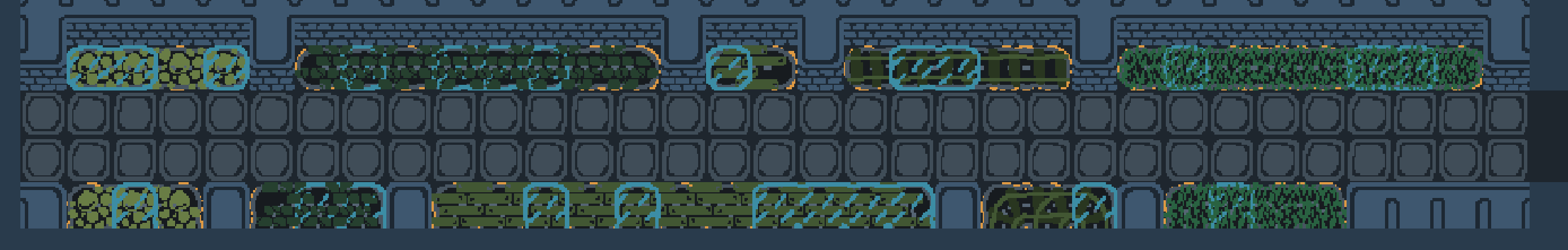}
  \end{center}
  \caption[]{The bottom five rows of the level. Unlike the two other main components of the level, this area can appear quite different depending on the particular instance of \textsc{pcp5-2} from which the level is being produced, but this is one example of how this section could appear.}\label{fig:bottom}  
\end{figure}

In the fourth row from the bottom, at the very left, add a wall. Then, for each character in \(\alpha_1\), working from left to right in that same fourth row, add a stack of a road object, a rubble object, and a lava object if the character is a 1 or an ice object if the character is a 0. Add another wall after this, then do the same for \(\alpha_2\) replacing rubble with brick, a wall, the same for \(\alpha_3\) replacing rubble with plank, a wall, the same for \(\alpha_4\) replacing rubble with track, a wall, and the same for \(\alpha_5\) replacing rubble with grass. Then, add walls from the end of this row all the way to the right of the level. Do the same on the bottom row with the \(\beta\) strings, except exchanging lava and ice for each other. Finally, extend the rows which are second and third from the bottom of the level to the right past the level border, adding the arbitrarily long two-tile-high space to the level specified in the description of \textsc{baba+2}. An example of how this construction might appear is visible in Figure~\ref{fig:bottom}. This completes the construction of the level.

\subsection{Proof of reduction}

We now demonstrate that the \textsc{baba+2} decision problem constructed above has the same answer as the \textsc{pcp5-2} instance on which it is based.

First, assume that there exists a solution to the instance of \textsc{pcp5-2}. In this case, there is some finite list of \(n \ge 1\) integers each between one and five \(i_1,i_2,\cdots,i_n\) such that \(\alpha_{i_1}\alpha_{i_2}\cdots \alpha_{i_n}=\beta_{i_1}\beta_{i_2}\cdots \beta_{i_n}\). In this case, the player can push up on the \(i_n\)'th \rl{is} word, then the \(i_{n-1}\)'th \rl{is}, and so on until they have pushed the \(i_1\)'th \rl{is}. This will replicate the tiles at the bottom of the level in reverse order to the \textsc{pcp5-2} solution. Since the first strings copied will end up the farthest to the right, this means that the third to bottom row of the level, when read left to right treating lava as 1 and ice as 0, will be the binary string \(\alpha_{i_1}\alpha_{i_2}\cdots\alpha_{i_n}\). Similarly, the binary string reached by reading the row second from the bottom with ice as 1 and lava as 0 will be \(\beta_{i_1}\beta_{i_2}\cdots \beta_{i_n}\), which is the same string by the definition of the sequence \(i_1,i_2,\cdots,i_n\). For these strings to be equal, there must be a lava object in the top row in each column where there is an ice object in the bottom row, and an ice object in the top row in any column where there is a lava object in the bottom row. Thus, each row will have the same number of objects, with exactly one ice and one lava object in each column. As a result, when the player pushes down on \rl{wall}, they can move down to eliminate all ice and lava objects and win the level. Thus, if the \textsc{pcp5-2} instance is a yes instance, then this \textsc{baba+2} instance is also a yes instance.

The only ways in which the player may interact with the level up until pushing \rl{wall} down are the five button gadgets where they may push \rl{is} up. There is no way for the player to do anything but push these five buttons in some order and then push \rl{wall} down to win the level, since the rule \rl{level without lava and without ice is win} is the only way to win the level, and the only way to remove the lava and ice tiles present at the beginning of the level is pushing \rl{wall} down. 

Thus, the only actions the player can legally take that could possibly result in winning the level are pushing up on the \rl{is} text objects some number of times (but at least once) in some order, followed by pushing down on \rl{wall}, then moving down or up once. Additionally, the moves taken between pressing the \rl{is} text up or pushing \rl{wall} down cannot affect anything in the level, so they can be ignored. Thus, if this level can be won, it is through some sequence of moves that can be completely described by the order in which various \rl{is}'s are pushed up.
 
If the level is won after such a sequence of moves, then on the turn when the player pushes \rl{wall} down, the third row from the bottom consists of some sequence of lava and ice tiles such that the second row from the bottom contains lava in any column the other row contains ice, it contains ice in any column in which the other row contains lava, and it is empty in any row in which the other row is empty. In any other configuration, pushing \rl{wall} down and moving in any direction does not result in the level being won. Thus, the sequence in which the player pushed the various \rl{is} text objects up resulted in a construction of a lava/ice string in the upper row that is exactly opposite in every position to the lava/ice string in the lower row. 

If we take the \(\alpha\) words in the reverse order that the player pushed the \rl{is} objects up (e.g. if the player first pushes the fourth \rl{is} up, then the fifth, we would take the string \(\alpha_5\alpha_4\)), we must construct a string which is the same string we would obtain from reading left to right in the row third from the bottom of the level, writing 1 for every lava object and 0 for every ice object. Similarly, carrying out the same process with the \(\beta\) words gives the same string we obtain from reading along the row second from the bottom of the level, writing a 0 for every lava object and a 1 for every ice object. However, we have seen that in each position where the lower row has an ice, and upper row has a lava, and vice versa. This means that the binary strings corresponding to each row must be the same, so we have found a solution to the instance of \textsc{pcp5-2}. Thus, the \textsc{pcp5-2} instance is a yes instance whenever \textsc{baba+2} is a yes instance.

Combining these results, we find that the \textsc{baba+2} instance is a yes instance if and only if the \textsc{pcp5-2} instance is a yes instance. The level's size is polynomial in the size of the \textsc{pcp5-2} instance and so the level can be constructed from the \textsc{pcp5-2} instance in polynomial time. As a result, this is a polynomial time reduction from \textsc{pcp5-2} to \textsc{baba+2}, so since \textsc{pcp5-2} is undecidable, \textsc{baba+2} is also undecidable.

\section{Conclusion}

We have demonstrated that the game \textit{Baba is You} is undecidable, though this proof required a generalization of the game with an arbitrarily long hallway adjoined to one side of the level. In general, some modification to the game is certainly necessary to reach this undecidability result, as the unmodified game is decidable. It is likely that other similarly small modifications to the rules of the game could also lead to this same undecidability result. In particular, removing the cap of the number of copies of an object which may be present on one tile should enable the construction of a counter machine for a proof along the same lines as the proof that Braid is undecidable given in \cite{hamilton2014braid}.

\bibliography{Baba_is_You_is_Undecidable}{}

\end{document}